\documentclass[12pt]{article}
\usepackage[utf8]{inputenc}
\usepackage[T1]{fontenc}
\usepackage[english]{babel}
\usepackage{amsmath,amsfonts,amssymb}
\usepackage{graphicx,caption}
\usepackage{geometry}
\usepackage{xcolor}
\usepackage{authblk}
\usepackage{lineno}
\usepackage{lscape}
\usepackage{natbib}
\usepackage{hyperref}
\hypersetup{
    bookmarks=true,         % show bookmarks bar?
    unicode=false,          % non-Latin characters in Acrobatâ_Ts bookmarks
    pdftoolbar=true,        % show Acrobatâ_Ts toolbar?
    pdfmenubar=true,        % show Acrobatâ_Ts menu?
    pdffitwindow=false,     % window fit to page when opened
    pdfstartview={FitH},    % fits the width of the page to the window
    pdftitle={My title},    % title
    pdfauthor={Author},     % author
    pdfsubject={Subject},   % subject of the document
    pdfcreator={Creator},   % creator of the document
    pdfproducer={Producer}, % producer of the document
    pdfkeywords={keyword1, key2, key3}, % list of keywords
    pdfnewwindow=true,      % links in new PDF window
    colorlinks=true,       % false: boxed links; true: colored links
    linkcolor=red,          % color of internal links (change box color with linkbordercolor)
    citecolor=blue,        % color of links to bibliography
    filecolor=cyan,         % color of file links
    urlcolor=magenta        % color of external links
}

\hypersetup{colorlinks=true,linkcolor=blue,urlcolor=blue,citecolor=blue}
\newcommand{\keywords}[1]{\bigskip\noindent\small{\bf\textit{Keywords--- }}{#1}}

\title{Dynamical environments of (486958) Arrokoth: \\ prior evolution and present state}
\author[1,2]{Ivan~I.~Shevchenko}
\author[3]{Jos\'e Lages}
\author[2]{Dmitrii~E.~Vavilov}
\author[3]{Guillaume Rollin}
\affil[1]{\small Saint Petersburg State University, 7/9 Universitetskaya nab., 199034 Saint Petersburg, Russia}
\affil[2]{\small Institute of Applied Astronomy, Russian Academy of Sciences, 191187 Saint~Petersburg, Russia}
\affil[3]{\small Institut UTINAM, CNRS, Universit\'e Bourgogne Franche-Comt\'e, Besan\c{c}on, France}

\date{\today}

\begin{document}

\maketitle
\begin{abstract}
We consider dynamical environments of (486958) Arrokoth, focusing
on both their present state and their long-term evolution,
starting from the KBO's formation. Both analytical (based on an
upgraded Kepler-map formalism) and numerical (based on massive
simulations and construction of stability diagrams in the 3D
setting of the problem) approaches to the problem are used. The
debris removal is due to either absorption by the KBO or by
leaving the Hill sphere; the interplay of these processes is
considered. The clearing mechanisms are explored, and the debris
removal timescales are estimated. We assess survival opportunities
for any debris orbiting around Arrokoth. The generic chaotization
of Arrokoth's circumbinary debris disk's inner zone and generic
cloudization of the disk's periphery, which is shown to be essential
in the general 3D case, naturally explains the current absence of
any debris in its vicinities.
\end{abstract}


\section{Introduction}

The Kuiper belt object (KBO) 2014~MU69, now called (486958)
Arrokoth, was the second (after Pluto) target object for the New
Horizons space mission. Even before the flyby, due to especial
observational campaigns \citep{S17,P17} it was classed as a
primordial contact binary (CB), presumably a typical KBO. The
flyby of New Horizons close to Arrokoth (January 1, 2019) showed
that, indeed, Arrokoth has a perfect contact-binary shape
\citep{S19LPI,C19LPI,P19LPI,Stern19Sci}, although its two
constituents are somewhat flattened \citep{Stern19Sci}. The ratio
of the constituents' masses turned out to be $\sim$1/3; this ratio
is rather typical for contact-binary cometary nuclei (see Table~1
in \citealt{LSR18}).

No satellites, moonlets, fragments, particles or any other debris
have been identified to be present in Arrokoth's vicinities,
although dedicated specialized surveys were performed from HST and
New Horizons \citep{K18,SSL19LPI,SSM20Sci,G19LPI}. The data
obtained during the New Horizons mission \citep{SSM20Sci} showed
that Arrokoth has no moonlets larger than 300~m in diameter (i.e.,
$\sim$1\% of Arrokoth's full size) throughout most of its Hill
sphere; it was also found that any rings around Arrokoth, if
present, are at least twice or thrice fainter in forward and
backward scattering than Jupiter's main ring.

The problem of emergence and survival of such kind of low-mass
material is important in two major respects.

First, cosmogonical: such material, if any, can be formed by
ejecta from the CB-forming collision \citep{U19LPI}; it may
represent remnants from a primordial swarm of solids
\citep{MK19LPI}; it may represent ejecta due to early out-gassing
\citep{T15AA,SL00JGR} or due to close encountering with other KBOs
\citep{N18AJ}. Any scenario of CBs formation in the Kuiper belt,
apart from explaining the occurrence of such slowly rotating
objects, should explain how the remnant debris are cleared away
\citep{U19LPI}.

Second, the problem is obviously important for planning any
survivable space missions that include close flybys.

In \cite{RSL21}, we explored properties of the long-term dynamics
of particles (moonlets, fragments, debris; either ordinary-matter
or dark-matter) around Arrokoth, as well as around similar
contact-binary objects potentially present in the Kuiper belt.
This was performed in the planar (2D) setting: it was assumed that
the motion of particles are planar and take place in the rotation
plane of Arrokoth, i.e., in the plane orthogonal to the angular
momentum vector of Arrokoth. The host dumbbell rotates in this
plane. Gravitational perturbations from the Sun, which could
invalidate the problem's planarity (because the dumbbell's
rotation plane is almost perpendicular to the ecliptic plane, see
data below), were not taken into account. In \cite{RSL21}, the
chaotic diffusion of particles inside the Hill sphere of Arrokoth
(or, generally, a similar object) was studied by means of
construction of appropriate stability diagrams and by application
of analytical approaches generally based on the Kepler map theory.

As it is well known, rotating CBs create zones of dynamical chaos
around them \citep{LSS17}. Any low-mass material orbiting in this
chaotic zone around a rotating dumbbell cannot survive: sooner or
later it either escapes from this zone, or fall on the host CB's
surface \citep{LSR18}; in this way, immediate vicinities of any
rotating CB are cleared. The chaotic zone formation in this case
is due to overlap of the orbit-spin resonances between the
orbiting particle and the rotating host dumbbell; any low-mass
material injected into the chaotic zone exhibits chaotic diffusion
in the eccentricity and other orbital elements and is sooner or
later removed \citep{LSR18}.

In fact, various types of non-uniformly shaped objects, including
contact-binary ones, were theoretically studied and identified to
create zones of orbital instability around themselves
\citep{mysen06,LSS17,madeira22}. Here we use the contact-binary
model as one straightforwardly relevant to the MU69 case.

Here we consider dynamical environments of (486958) Arrokoth,
focusing on both their present state and their long-term
evolution, starting from the KBO's formation. By the dynamical
environments of Arrokoth we imply the modes of motion of any
actual or possible populations of passively gravitating particles
inside Arrokoth's Hill sphere.

The work presented here is a ``3D continuation'' of the work by
\cite{RSL21}: from the planar setting of the problem we proceed to
the 3D one. Both analytical (based on an upgraded Kepler-map
formalism) and numerical (based on massive simulations and
construction of stability diagrams) approaches to the problem are
used. We assess survival opportunities for any debris inside
Arrokoth's Hill sphere. The debris removal is due to either
absorption by the KBO or by leaving the Hill sphere; the interplay
of these processes is considered. The clearing mechanisms are
explored, and the debris removal timescales are estimated in the
3D setting of the problem, taking into account the actual
dynamical parameters of Arrokoth (in particular, its angular
momentum vector orientation in space) and gravitational
perturbations from the Sun.

\section{The problem setting and numerical simulations}
\label{sec_numsim3}

\subsection{The 2D problem}

In the 2D setting of the problem, explored in \cite{RSL21}, the
chosen inertial Cartesian coordinate system has the origin at the
CB's center of mass, and the equations of motion of a passively
gravitating particle with coordinates $x$, $y$ are given by

\begin{equation}
    \begin{array}{ccl}
        \dot{x}&=&v_x, \\
        \dot{y}&=&v_y, \\
        \dot{v}_x&=&-{{{\it m_2}\,\left(x-{\it x_2}\right)}\over{\left(\left(y-
                {\it y_2}\right)^2+\left(x-{\it x_2}\right)^2\right)^{{{3}/{2}}}
        }}-{{{\it m_1}\,\left(x-{\it x_1}\right)}\over{\left(\left(y-
                {\it y_1}\right)^2+\left(x-{\it x_1}\right)^2\right)^{{{3}/{2}}}
        }},\\
        \dot{v}_y&=&-{{{\it m_2}\,\left(y-{\it y_2}\right)}\over{\left(\left(y-
                {\it y_2}\right)^2+\left(x-{\it x_2}\right)^2\right)^{{{3}/{2}}}
        }}-{{{\it m_1}\,\left(y-{\it y_1}\right)}\over{\left(\left(y-
                {\it y_1}\right)^2+\left(x-{\it x_1}\right)^2\right)^{{{3}/{2}}}
        }},
    \end{array}
    \label{motion1}
\end{equation}

\noindent where the coordinates $x_1$, $y_1$ and $x_2$, $y_2$
designate the Cartesian locations of the mass centers of the CB's
constituents with masses $m_1$ and $m_2$, respectively:

\begin{equation}
    \begin{array}{ccl}
        x_1&=&\mu\cos \omega t , \\
        y_1&=&\mu\sin \omega t , \\
        x_2&=&(\mu-1)\cos \omega t ,\\
        y_2&=&(\mu-1)\sin \omega t ,
    \end{array}
    \label{motion2}
\end{equation}

\noindent where $\omega$ is the CB's angular rotation frequency in
units of the CB's critical rotation rate, corresponding to its
centrifugal disintegration. Note that, if one sets $\omega = 1$,
then the equations are nothing but the usual equations of motion
in the planar circular restricted three-body problem. We define
the mass parameter $\mu = m_2/(m_1+m_2)$, where $m_2 \leq m_1$.
The distance between the mass centers of $m_1$ and $m_2$ is
constant and is set to unity, $d=1$; this defines the length unit.
We set $\mathcal{G}(m_1+m_2)=1$, where $\mathcal{G}$ is the
gravitational constant. The critical rotation rate (the angular
Keplerian velocity of the binary, if it were unbound) is $\omega_0
= \left[ \mathcal{G}(m_1+m_2)/d^3 \right]^{1/2} = 1$. Note that
the typical rotation rates $\omega$ of KBOs range from one fifth
to one $\omega_0$, i.e., the rotation periods range from one to
five, if expressed in critical periods; see KBO lightcurve data in
\cite{TNO14}.

In \cite{RSL21}, we used the physical and dynamical data for
Arrokoth as obtained during the New Horizons flyby
\citep{S19LPI,C19LPI,P19LPI,Stern19Sci}. To describe the
immediate dynamical environments of Arrokoth, we constructed the
stability chart, namely, the Lyapunov characteristic exponent
(LCE) diagram, in the ``pericentric distance -- eccentricity''
plane $q$--$e$, on a fine grid of initial conditions. The diagram
is shown in Figure~2 in \cite{RSL21}, its most prominent feature
is the fractal ``ragged'' border between the chaotic and regular
zones. The ragged border is formed by orbit-spin resonances
between an orbiting particle and the rotating Arrokoth. The most
prominent bands of chaos are formed by the $p/1$ orbit-spin
resonances, corresponding to integer ratios of the particle's
orbital period and Arrokoth's rotation period.

\subsection{The 3D problem}

\begin{figure}[!t]
\centering
\includegraphics[width=0.8\linewidth]{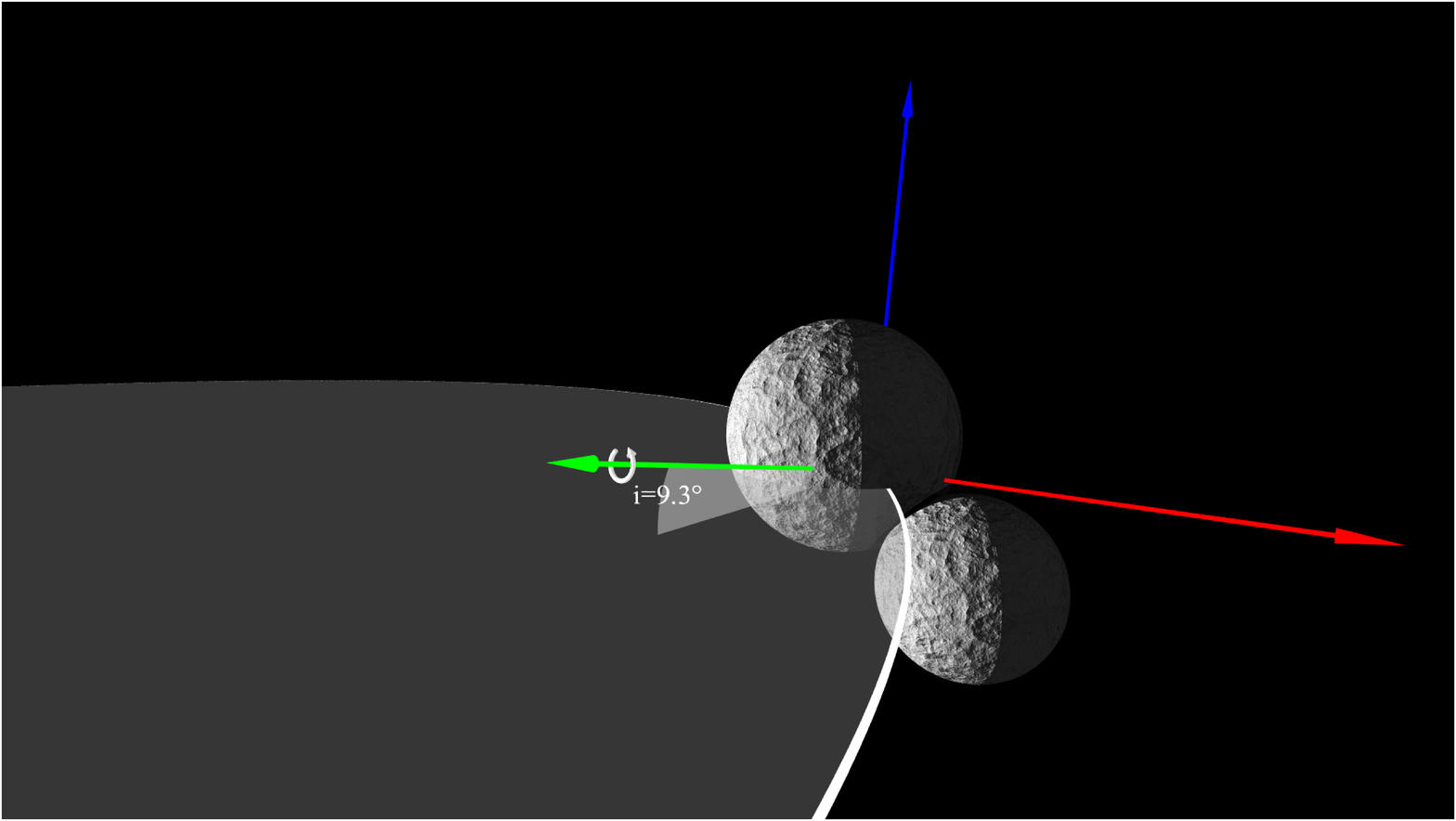}
\caption{The object scheme and the adopted Arrokoth-centered
coordinate system. The axis $z$ (shown in green) of the coordinate
system coincides with Arrokoth's rotation axis. The $x$ and $y$
axes are in red and blue, respectively. Arrokoth's heliocentric
orbit and orbital plane are in white and grey, respectively. The
sizes (diameters) of the lobes are $\approx$20~km and
$\approx$14~km, and the mean distance from the Sun is
$\approx$45~AU. } \label{Fig1}
\end{figure}

In the present work, we generalize the dynamical model and explore
the dynamical environments of Arrokoth in the 3D setting of the
problem, taking into account the actual 3D dynamical parameters of
Arrokoth (in particular, its angular momentum vector orientation
in space) and gravitational perturbations from the Sun. In
Figure~\ref{Fig1}, the object scheme and the adopted coordinate
system are graphically presented.

To describe the dynamical environments of Arrokoth, we construct
stability charts in the $q$--$\omega$ (pericentric distance ---
angular velocity) plane of initial conditions, now in the 3D
setting. We choose an inertial Cartesian coordinate system with
the origin at the Arrokoth's mass center. In addition to the
gravitational effect of Arrokoth, we take into account
perturbations from the distant Sun. The gravitational potential of
Arrokoth is considered as a potential of two mass points, the
distance between which is constant in time. The $z$-axis of the
coordinate system coincides with the spin axis of Arrokoth. The
$x$ and $y$ axes complete the $(x,y,z)$ set to an orthogonal one;
see Figure~\ref{Fig1}.

In the general 3D setting, the equations of motion of a passively
gravitating particle with radius-vector $\mathbf{r}$ and velocity
$\mathbf{v}$ are given by

\begin{equation}
    \begin{array}{l}
        \dot{\mathbf{r}} = \mathbf{v} , \\
        \dot{\mathbf{v}} = - \mathcal{G} m_1\frac{\mathbf{r}-\mathbf{r_1} }
        {|\mathbf{r}-\mathbf{r_1}|^3} - \mathcal{G} m_2\frac{\mathbf{r}-\mathbf{r_2}
        } {|\mathbf{r}-\mathbf{r_2}|^3} + \mathbf{A}_\mathrm{pert} ,
    \end{array}
    \label{eq:motion}
\end{equation}
\noindent where
\begin{equation}
    \mathbf{A}_\mathrm{pert} =
    - \mathcal{G} m_{\odot}\frac{\mathbf{r}-\mathbf{r_{\odot}} }
    {|\mathbf{r}-\mathbf{r_{\odot}}|^3}
    - \mathcal{G} m_{\odot}\frac{\mathbf{r_{\odot}} } {|\mathbf{r_{\odot}}|^3}
    \label{eq:motion1}
\end{equation}

\noindent is  the Solar perturbation; $\mathbf{r_1}$,
$\mathbf{r_2}$ and $\mathbf{r_{\odot}}$ are the radius-vectors of
Arrokoth's first and second lobes and the Sun, respectively;
$m_1$, $m_2$ and $m_{\odot}$ are their masses.

We assume that the gravitational influence of the Sun on the spin
axis orientation and the rotation rate $\omega$ of Arrokoth is
negligible. The locations of the centers of Arrokoth's two lobes
are given by

\begin{equation}
    \begin{array}{l}
        \mathbf{r_1} = (r_1 \cos \omega t , r_1 \sin \omega t , 0) , \\
        \mathbf{r_2} = (r_2 \sin \omega t , r_2 \cos \omega t , 0) ,
    \end{array}
    \label{eq:motion2}
\end{equation}

\noindent where $r_1$ and $r_2$ are the distances from the CB's
mass center to the mass centers of its lobes.

For a host dumbbell, consisting of two round lobes, as Arrokoth
is, the dynamical model given by
Equations~(\ref{eq:motion})--(\ref{eq:motion2}) is obviously
suitable. In vicinities close to the object's surface, the
irregularities of the latter may affect the orbiting particles,
especially their accretion onto the object's surface. However note
that any particles with the orbital pericentric distance $q
\lesssim 2 d$ (where $d$ is the CB's size) are absorbed by
Arrokoth almost immediately \citep{RSL21}.

Equations~(\ref{eq:motion})--(\ref{eq:motion2}) were integrated by
the 15th order Everhart integrator~\cite{E85IAU}. The local
relative error tolerance was set to $10^{-15}$.

In view of the general scenarios of formation of CB KBOs
\citep{U19LPI,MK19LPI}, we assume, as in \cite{RSL21}, that the
post-formation phase of Arrokoth's evolution starts with the
particles initially residing in a disk-like structure formed
around the merged CB. In case of Arrokoth, the circumbinary
chaotic zone in the disk may extend up to radii $\simeq 6 d$
\citep{RSL21}. We sampled 400 angular velocity values from 0.01 to
15.78 rad per day; this corresponds to periods from 628.32~d to
9~hr 13~min. The pericentric distances were sampled from 19~km to
130~km with a step of 0.2~km. The parameters of Arrokoth were set
as following:

\begin{enumerate}

    \item The ''dumbbell size'' of Arrokoth (the distance between the
    mass centers of its two lobes): $d = 16.1$ km. Radii of the
    components are $R_1 = 10$~km and $R_2 = 7$~km; see Figure~1 in
    \cite{C19LPI}.
    The rotation period of the dumbbell is 15.9~h \citep{K22JGR}.

    \item The masses (assuming the density value $\rho = 0.5$ g/cm$^3$,
    typical
    for cometary nuclei; see \citealt{T13Ic,J16Ic}) of the lobes are
    $m_1 = 1.70 \cdot 10^{18}$~g and $m_2 = 6.67 \cdot 10^{17}$~g
    (equal to $8.55 \cdot 10^{-16}$ and $3.35 \cdot 10^{-16}$
    in the Solar mass units); therefore, the mass parameter
    $\mu_\mathrm{A} = m_2/(m_1 + m_2) \approx 0.282$.

    \item The Arrokoth's orbit around the Sun is considered, for our
    purposes, as circular, as its eccentricity is small
    ($e_\mathrm{A} = 0.044$);
    for the semimajor axis value we set
    $a_\mathrm{A} = 44.67$~AU \citep{JPL22};
    therefore, the period of motion around the Sun is
    298.6~yr.

    \item The Arrokoth's obliquity (the angle between the spin axis of
    Arrokoth and the normal to its orbital plane) is set to
    $99.3^{\circ}$, in accordance with data in \cite{P19EPSC}.

    \item The initial eccentricity of the disk particles is set to
    zero. The particles start orbiting in the Arrokoth's rotation
    plane.

    \item The computing maximum time is set to 1000~yr, covering
    $\approx$3.3 Arrokoth's revolutions around the Sun.

\end{enumerate}

The particle's orbit is integrated until the computing maximum
time is reached or until the particle collides with the asteroid
or leaves Arrokoth's Hill sphere, whose radius is
$\approx$50000~km, equal to $\approx 7.5 \cdot 10^{-6}$ in units
of the radius of Arrokoth's orbit around the Sun.

The adopted here maximum computing time (1000~yr) is by far
sufficient for our current purposes, because the total clearing of
the circumbinary chaotic zone takes place on much shorter (by
orders of magnitude) timescales, as we find out further on. What
is more, the typical orbital periods of particles inside the
chaotic zone ($\sim$ hours or days) are also by far shorter.

Concerning Arrokoth's orbital eccentricity, it is very small $(e
\approx 0.04)$, therefore, we do not expect any visible change in
the integration results if the eccentricity is taken into account.
Further on, we discuss this issue in more detail.

\begin{figure}
\includegraphics[width=\linewidth]{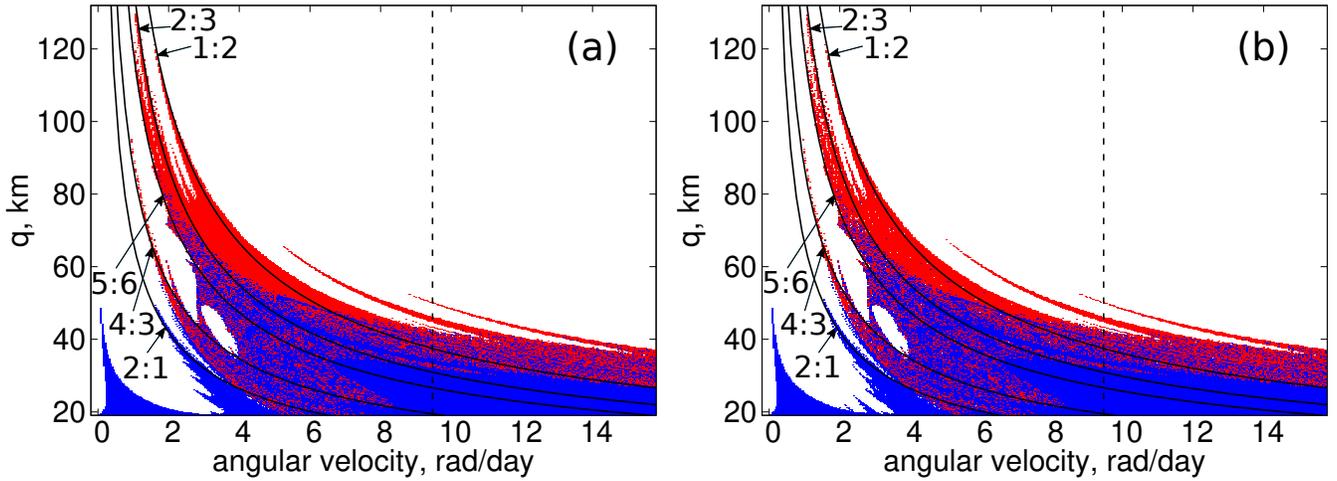}
%\bigskip
\caption{The stability diagrams. (a)~Solar perturbations are not
taken into account; (b)~Solar perturbations are taken into
account. Blue domain: particles colliding with the host CB
surface; red domain: particles escaping the CB's Hill sphere;
white domain: none of these events taking place in the course of
the maximum time interval (1000~yr) of integration. Black solid
curves: the locations of several orbit-spin resonances of the
orbiting particle with the rotating CB. The dashed vertical line
corresponds to the Arrokoth's current rotation rate, as determined
by~\cite{P19EPSC}.}
\label{img:Stab_diag}
\end{figure}

In the resulting diagrams in Figs.~\ref{img:Stab_diag}, we
represent graphically the extents of the circum-CB chaotic zone,
as determined in our computations, with the Solar perturbations
not taken into account (Fig.~\ref{img:Stab_diag}a) and taken into
account (Fig.~\ref{img:Stab_diag}b).

The diagrams are defined in the ``particle's initial pericentric
distance -- CB rotation rate'' ($q$--$\omega$) frame. In the
diagrams, blue color means that the particle collides with the
CB's surface, red color means that it escapes the CB's Hill
sphere, and white color that none of such events have happened
during the maximum (1000~yr) interval of integration. Comparing
the diagrams in Fig.~\ref{img:Stab_diag}a and
\ref{img:Stab_diag}b, one finds that in the both considered cases
(without and with Solar perturbations) the results are almost
identical: the chaos borders are almost the same.

\begin{figure}
\includegraphics[width=\linewidth]{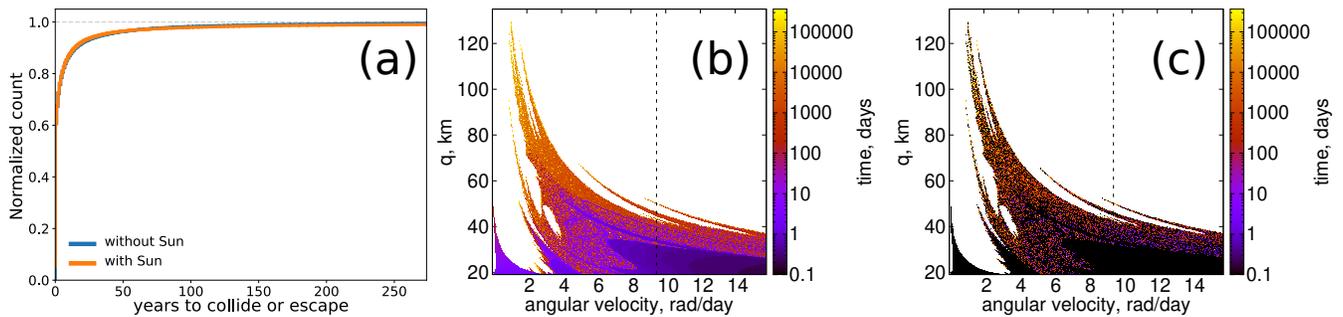}
\caption{(a)~The cumulative distribution for the survival time of
particles, in the cases without (blue) and with (yellow) Solar
perturbations taken into account. (b)~The time to collide or
escape as a function of the CB's angular velocity (rotation rate)
and initial perihelion distance. The dashed vertical line marks
the Arrokoth's current angular rotation rate, as determined
by~\cite{P19EPSC}. (c)~The survival time difference between
the cases with and without the Sun.}
\label{img:Time_to_collide}
\end{figure}

In Fig.~\ref{img:Time_to_collide}a, the cumulative distribution is
shown for the time required for a particle to collide with
Arrokoth or escape Arrokoth's Hill sphere. One finds that almost
all of the particles collide or escape in 100~yr. A half of the
particles has lifetime less than 50~d. No significant differences
in results are observed between the two considered cases (with or
without the Solar perturbations taken into account). In
Fig.~\ref{img:Time_to_collide}b, the time needed for a particle to
collide with Arrokoth or escape Arrokoth's Hill sphere as a
function of Arrokoth's initial perihelion distance and rotation
rate is presented. Since there are no significant differences, we
present solely the case with the Solar perturbations taken into
account. Fig.~\ref{img:Time_to_collide}c shows the difference
in lifetime between the cases with and without the Sun. One may
see that the difference is typically much less than the lifetime
values themselves (note that the latter are especially large in
the upper-left corner of the diagram). However, this inference
concerns Arrokoth's close vicinities covered by the diagram. For
much larger orbits, the Solar perturbations become dominant, as
discussed further on.

For Arrokoth's current rotational period, the particles that are
closer in orbital radius to Arrokoth's mass center than 32~km are
observed collide with Arrokoth in several days. This is in accord
with results by \cite{2022Ap&SS.367...38A}. The particles that
move initially farther in orbital radius, may collide or escape
only if they follow orbits close to orbit-spin resonances
(resonances between particle's orbital period and Arrokoth's
rotation period); therefore, the survival time can generally be
much greater.

\section{Dispersal of matter around Arrokoth}
\label{sect_DmaA}

The diffusion coefficient is as usual defined as the mean-square
spread in a selected variable (say, energy $E$), per time unit:
\begin{equation}
    D \equiv \lim_{t \to \infty} \frac{\langle (E(t) - E_0)^2
        \rangle}{t} , \label{defD}
\end{equation}
\noindent where $t$ is time, and the angular brackets denote averaging
over a set of starting values of $E$; see \cite{Meiss92}.

As determined in the 2D problem setting \citep{RSL21}, for a CB
like Arrokoth (with the mass ratio $\mu \sim 0.1$--0.3) the
characteristic timescale of the diffusion ($\sim 1/D$) in the
immediate circumbinary chaotic zone can be as small as $\sim$10
times CB's rotation period. This means that the clearing of the
chaotic zone is essentially instantaneous. Although this estimate
of the transport time was performed in the diffusional
approximation, its smallness means that, in fact, the transport is
not diffusional, but ballistic \citep{RSL21}.

In \cite{RSL21}, the ballistic transport character was shown
independently by calculating the amplitude of the kick function
for the generalized Kepler map; on the generalized Kepler map
theory see \cite{LSS17,LSR18}. The kick function in the energy $E$
is given by
\begin{equation}
    \Delta E\left(\mu,q,\omega,\phi\right) \simeq
    W_1\left(\mu,q,\omega\right)\sin\left(\phi\right)+
    W_2\left(\mu,q,\omega\right)\sin\left(2\phi\right) ,
    \label{deltaEall}
\end{equation}
\noindent where $\phi$ is the CB's phase angle when the orbiting
particle is at pericenter, and
\begin{equation}\label{W1}
    W_1\left(\mu,q,\omega\right)\simeq \mu\nu(\nu-\mu)2^{1/4}
    \pi^{1/2} \omega^{5/2} q^{-1/4} \exp \left( - \frac{2^{3/2}}{3}
    \omega q^{3/2} \right) ,
\end{equation}
\begin{equation}\label{W2}
    W_2\left(\mu,q,\omega\right)\simeq-\mu\nu2^{15/4} \pi^{1/2}
    \omega^{5/2}q^{3/4}\exp \left( -
    \frac{2^{5/2}}{3} \omega q^{3/2} \right) ;
\end{equation}
\noindent and, by definition, $\nu = 1 - \mu$.

One may see that, at $\mu \sim 1/3$, $\omega \sim 1$ and $q
\sim$2--3 one has $W_1 \sim W_2 \sim 1$. Therefore, the
single-kick energy variation is $\sim 1$; this means that, indeed,
any orbiting particle that is initially placed in the CB's chaotic
zone can be ejected from this zone in a few orbital revolutions
\citep{RSL21}. This conclusion, obtained in the 2D setting of the
problem, is not expected to be subject to change in the 3D
setting. However, the statistical behaviour of particles in the
removal time distribution tail can be different. We explore this
possibility in the following Sections.

\section{Cloudization of low-mass matter in the outer orbital zone}
\label{sect_cloud}

According to \cite{RSL21}, the originally formed low-mass matter
cocoon inside Arrokoth's Hill sphere could have suffered $\sim
10$--100 dispersal events since Arrokoth's formation epoch;
therefore, one may expect that Arrokoth's Hill sphere (as well as
Hill spheres of similar KBOs) nowadays is empty.

Recall that the radius $R_\mathrm{H}$ of the secondary body's (say
with mass $m_\mathrm{A}$) Hill sphere, in units of semimajor axis
of the body's orbit around the primary mass $m_0$, is
\begin{equation}
    R_\mathrm{H} = \left( \frac{\mu}{3} \right)^{1/3} ,
    \label{Hill_sma}
\end{equation}
\noindent where $\mu = m_\mathrm{A}/(m_0 + m_\mathrm{A})$, as
usual, is the mass parameter of the system (here, the
Sun--Arrokoth one). The orbit of Arrokoth's any moonlet should
lie within Arrokoth's Hill sphere. This implies the inequality
$a(1 + e) \lesssim R_\mathrm{H}$, for the semimajor axis $a$
and eccentricity $e$ of the moonlet.

Given the ``dumbbell size'' of Arrokoth $d = 16$~km, it is
straightforward to estimate \citep{RSL21} that the chaotic
clearing zone around Arrokoth may have radius of at most
$\sim$100~km,\footnote{Note that the $\sim$100~km limit is in
accord with the stability diagrams in Figs.~\ref{img:Stab_diag}a
and \ref{img:Stab_diag}b.} an order of magnitude less than the New
Horizons flyby distance ($\sim$3500 km) and three orders of
magnitude less than Arrokoth's Hill radius ($\sim 5 \cdot
10^4$~km).

How the low-mass matter cocoon inside the Hill sphere may emerge?
As revealed in \cite{T93}, in any star's planetary system, there
exists a critical semimajor axis at which the chaotic
highly-eccentric motion inside the host star's Hill sphere
transforms from the diffusion in energy at approximately constant
$q$ to the diffusion in angular momentum at constant energy. This
transition takes place where the torque from the Galactic tide
starts to dominate. The freezing in energy (equivalently, in
semimajor axis) forms an effective barrier for the escape process;
thus it is a necessary constituent for the formation of a cocoon
of matter inside the Hill sphere.

In the dynamical environments of a contact-binary KBO, quite
analogously, the critical radius, at which the diffusion in energy
(with the pericentric distance $q$ constant) is stopped and the
diffusion in angular momentum (with the semimajor axis $a$
constant) starts, can be estimated by equating the frequency of
circumbinary orbital precession to the frequency of the
Lidov--Kozai (LK) oscillations. The latter arise due to
perturbations from a distant perturber \citep{L61I,L62PSS,K62AJ},
the Sun in our case. The criteria for the Lidov--Kozai effect
suppression are described in \cite[section~3.3]{S17LKE}. Note
that, at the edge of escape of a particle from Arrokoth's Hill
sphere, its orbital period around Arrokoth is $\sim$500~yr, which
is greater (by a factor of two) than Arrokoth's heliocentric
orbital period (which is $\approx$300~yr) and is much greater (by
five orders of magnitude) than Arrokoth's rotation period (which
is $\approx$16~h); thus allowing one to use averaged equations of
motion.

Consider a close binary orbiting the main central mass $m_0$ (the
Sun in our case, $m_0 = m_{\odot}$). Thus, we have the
primary binary (the Sun--Arrokoth in our case) and the
secondary very compact binary (Arrokoth's two lobes) with
masses $m_1 > m_2$; see Figure~\ref{Fig1} for a general scheme.
The semimajor axes of the primary and secondary binaries are
designated henceforth $a_1$ and $a_2$, respectively; and their
eccentricities are $e_1$ and $e_2$.

\cite{MPG17} derived the following formula for estimating the
critical semimajor axis value $a_\mathrm{crit}$ (below which the
Lidov--Kozai effect is suppressed) for a particle orbiting
around the secondary binary in such an orbital system:
\begin{equation}
    a_\mathrm{crit} = \left[ \frac{3 m_1 m_2 \left(1 - e_1^2
        \right)^{3/2}}{8 m_0 (m_1 + m_2) \left( 1 - e^2 \right)^2 }
    \left| 5 \cos^2 i - 1 \right| \right]^{1/5} \left( a_1^3 a_2^2
    \right)^{1/5} , \label{acrit_bin}
\end{equation}
where $e$ and $i$ are the eccentricity and inclination of the
particle's orbit (here the inclination is referred to the orbital
plane of the secondary binary), all other quantities are defined
as just above, and $e_2$ among them is set to zero.

For the Sun--Arrokoth system one has $a_1 \gg a_2$ by many
orders of magnitude, whereas the mass of Arrokoth is by many
orders of magnitude smaller than the mass of the Sun,
$m_\mathrm{A} = m_1 + m_2 \ll m_0$. In our case, the secondary
binary is contact, therefore, it is circular, i.e., $e_2 = 0$, as
already set above.\footnote{Besides, we assume that the
rotation rate of the contact binary is critical, being equal to
the orbital velocity of the lobes around Arrokoth's mass center,
if they were physically unbound.} For a system with $e_1 \approx
0$ and $i \approx 0^\circ$ (as the ``Sun--Arrokoth--disk
particle'' initial configuration is) one may render
formula~(\ref{acrit_bin}) in the form
\begin{equation}
    a_\mathrm{crit} \approx \left[ \frac{3 \mu_1 \mu_2 a_1^3 a_2^2}{2
        \left( 1 - e^2 \right)^2 } \right]^{1/5} , \label{acrbc}
\end{equation}
where $\mu_1 = m_\mathrm{A}/(m_0 + m_\mathrm{A}) =
m_\mathrm{A}/m_0$ and $\mu_2 = m_2/(m_1 + m_2)= m_2/m_\mathrm{A}$
are, respectively, the mass parameters of the primary and
secondary binaries in our system;
$m_0$, $m_1$, and $m_2$ are, respectively, the masses of the Sun
and Arrokoth's lobes ($m_1 > m_2$); $a_1$ and $a_2$ are,
respectively, Arrokoth's orbital semimajor axis (of the orbit
around the Sun) and Arrokoth's dumbbell size; $e_1 =0.04 \simeq 0$
is the eccentricity of Arrokoth's orbit around the Sun.

The obliquity of the secondary binary's orbital plane (the
dumbbell's rotation plane, in the given case) with respect to the
primary binary's orbit is assumed to be high enough, so that the
Lidov--Kozai effect for particles orbiting around Arrokoth in
the plane of its rotation is potentially present (at $a >
a_\mathrm{crit}$). Note that the obliquity of Arrokoth's
rotation plane is indeed high, $\sim 99^\circ$ \citep{P19EPSC},
as needed.

For the necessary quantities to substitute in
Equation~(\ref{acrbc}), one has: $\mu_1 \approx 1.19 \cdot
10^{-15}$, $\mu_2 \approx 0.28$, $a_1 = 44.54$~AU $\approx 6.68
\cdot 10^{14}$~cm, $a_2 = d \approx 1.6 \cdot 10^6$~cm, and we
formally set $e=0$. Substituting the values of all these
quantities in Equation~(\ref{acrbc}), one finds $a_\mathrm{crit}
\simeq 2100$~km. Due to the specific dependence of
$a_\mathrm{crit}$ on $e$, the $a_\mathrm{crit}$ value is by the
order of magnitude the same for circular and eccentric
orbits, even at large $e$ values. As soon as $R_\mathrm{H} \simeq
10^5$~km, we find that $a_\mathrm{crit} \ll R_\mathrm{H}$.

One may thus be confident that the particles' random walk in
energy is effectively frozen already deep inside the Hill sphere,
and the low-mass matter cocoon may indeed be formed by
particles leaving the circumbinary chaotic zone, as well as
by any low-mass material initially residing in regular orbits in
the outer parts of the circumbinary disk.

Such material can be effectively ``cloudized,'' i.e.,
converted from the planar disk to a non-planar, approximately
spherically-symmetric cloud. Indeed, at $a > a_\mathrm{crit}$, the
Lidov--Kozai effect is not suppressed; therefore, the
highly-eccentric particles may suffer the LK-oscillations in their
eccentricity and inclination.
Let us approximately estimate the cloudization timescale.

As above, we assume the CB to move in a circular ($e_1 = 0$)
orbit of radius $a_1$ around the Sun. For the Keplerian orbital
elements of a particle orbiting around the CB (in the plane
of its rotation) we take, as usual, the semimajor axis,
eccentricity, inclination, argument of pericenter, longitude of
ascending node, and mean anomaly, denoted by $a$, $e$, $i$,
$\omega$, $\Omega$, and $M$, respectively.

In any two-body problem, the Delaunay variables are defined as
(see, e.g., \citealt{Morbi02,S17LKE})
\begin{equation}
    \begin{array}{lclcl}
        L &=& [ \mathcal{G} (m' + m'') a ]^{1/2} , \quad &l& = M , \\
        G &=& L ( 1-e^2 )^{1/2} , \quad &g& = \omega , \\
        H &=& G \cos i , \quad &h& = \Omega,  \\
        \label{var_D}
    \end{array}
\end{equation}
\noindent where $m'$ and $m''$ are the two masses, and
$\mathcal{G}$ is the gravitational constant. The variables $L$ and
$l$, $G$ and $g$, $H$ and $h$ form three pairs of conjugate
canonical variables. $L$ is a function of the semimajor axis
solely and thus can be expressed through the orbital energy; $G$
is the absolute value of the reduced (per unit of mass) angular
momentum; and $H$ is the reduced angular momentum vector's
vertical component. Therefore, $G$ is the specific angular
momentum, and its time derivative $\dot G$ is the torque (per unit
of mass).

In the presence of a perturbing third body in an outer orbit,
the equation for the Lidov--Kozai evolution of the angular
momentum $G$ of the inner binary is given by
\begin{equation} \dot G = - \frac{15 \mathcal{G} m_\mathrm{pert}
        a^2}{8a_\mathrm{pert}^3} e^2 \sin^2 i \sin 2\omega \label{dotG}
\end{equation}
\noindent \citep[Equation~3.26]{S17LKE}, where $m_\mathrm{pert}$
is the mass of the perturber. In our problem, the perturber
is the Sun, therefore, $m_\mathrm{pert} = m_0$, $a_\mathrm{pert} =
a_1$. Here Arrokoth is considered as a single gravitating point
(because the particle's orbit is assumed to be large enough), and
the particle's orbital inclination is referred to the primary
binary's orbital plane.

At present, Arrokoth's rotation plane is inclined by $\approx
99^\circ$ with respect to the ecliptic plane \citep{P19EPSC},
i.e., the obliquity is extremal. Therefore, we set $\sin i \simeq
1$ here. Also setting $e \simeq 1$ (since our particles are highly
eccentric) and substituting $| \sin 2\omega |$ by its averaged
(over an ensemble of particles) value $2 / \pi$, we obtain
\begin{equation}
    | \langle \dot G \rangle | = \frac{15 \mathcal{G} m_\mathrm{pert}
        a^2}{4 \pi a_\mathrm{pert}^3} .
    \label{avdotG}
\end{equation}
Let us define the characteristic timescale for the Lidov--Kozai
evolution as the time needed for $G$ to change by of order of
itself. This is just the characteristic time needed to convert the
initial ring (with radius $a$) of particles to a spherical cloud,
i.e., to make the initial planar distribution 3D-isotropic; we
designate this time $T_\mathrm{3D}$.

For a highly-eccentric passively gravitating particle, the reduced
angular momentum is given by
\begin{equation}
    G = \left[ \mathcal{G} m_\mathrm{CB} a ( 1-e^2 ) \right]^{1/2}
    \simeq ( 2 \mathcal{G} m_\mathrm{CB} q )^{1/2}
\end{equation}
(see Equations~(\ref{var_D})), where $m_\mathrm{CB}$ is the
CB's mass (the mass of Arrokoth, in our case), and $a$, $e$, $q$
are the test particle's semimajor axis, eccentricity and
pericentric distance, respectively. Equating $| \langle \dot G
\rangle | \simeq G / T_\mathrm{3D}$, one obtains
\begin{equation} T_\mathrm{3D} \simeq
    \frac{2^{5/2} \pi}{15} \, \frac{m_\mathrm{CB}^{1/2}
        a_\mathrm{pert}^3}{\mathcal{G}^{1/2} m_\mathrm{pert}} \,
    \frac{q^{1/2}}{a^2} . \label{T3D}
\end{equation}
Substituting $m_\mathrm{CB} = m_\mathrm{Arrokoth} \simeq 2.37
\cdot 10^{18}$~g, $m_\mathrm{pert} = m_0 = 1.989 \cdot
10^{33}$~g, $a_\mathrm{pert} = a_1 = 44.54$~AU$\approx 6.68 \cdot
10^{14}$~cm, $q =$(2--5)~$d_\mathrm{Arrokoth} =$(3.2--8.0)$\cdot
10^6$~cm, $a = R_\mathrm{H} \sim 10^{10}$~cm, one has:
$T_\mathrm{3D} \sim 0.6$--0.9~yr.

Given that, for a particle orbiting around Arrokoth with the
semimajor axis $a \sim R_\mathrm{H} \sim 10^5$~km, the orbital
period is $\sim 500$~yr, we see that the conversion process of the
periphery of the initial disk of particles to a spherical cloud,
if assessed in the orbital timescale, should be relatively
rapid. In what concerns the disk's inner zone, most of the
particles escape or are absorbed by Arrokoth in just a few
revolutions around the CB.

\begin{figure}
\begin{center}
\includegraphics[width=1.0\linewidth]{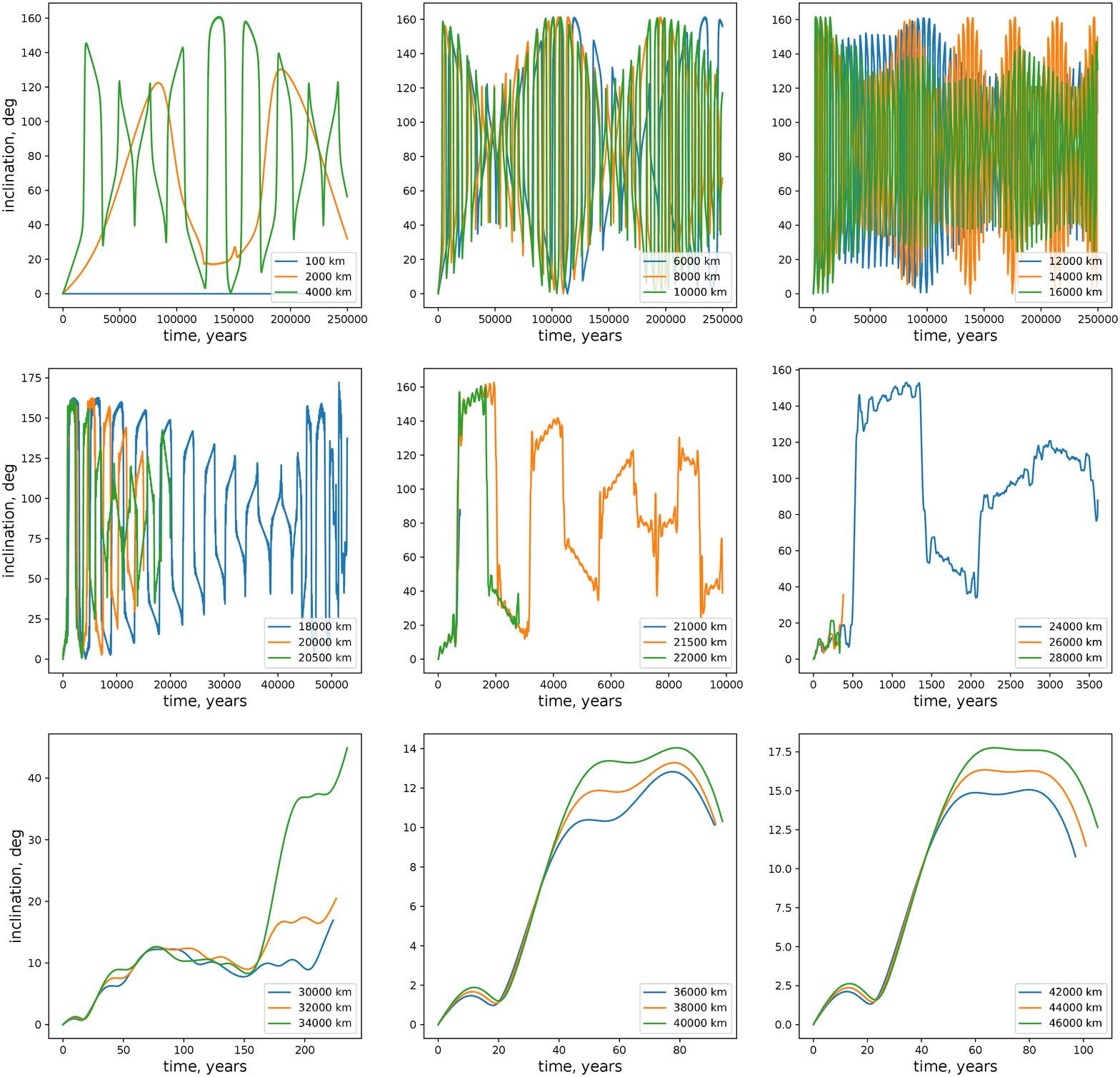}
\end{center}
        \caption{Orbital inclination of a particle as a function of time
        at various initial orbital semimajor axes.
        Initially the particle is placed in a circular orbit in the plane
        orthogonal to Arrokoth's spin axis.}
        \label{fig4}
\end{figure}

In Fig.~\ref{fig4}, the evolution of the orbital inclination
of particles with time is illustrated, as observed in our
numerical experiments. The particles were set to initially have
circular orbits with various values of the radius, in the plane
orthogonal to the spin axis of Arrokoth. In the Figure, the
curves of different colours (blue, orange and green) correspond to
three different initial values of the semimajor axis; these values
are indicated at each panel. The integration of motion is
performed in the same way as described in
Section~\ref{sec_numsim3}, but with the time limit of 250000~yr;
the CB's rotation rate is set equal to the current one. The
integrated orbits demonstrate that the particles with semimajor
axes greater than 16000~km do not survive on times greater than
50000~yr. The particles with smaller semimajor axes are mostly
long-lived.

\begin{figure}
\begin{center}
\includegraphics[width=1.0\linewidth]{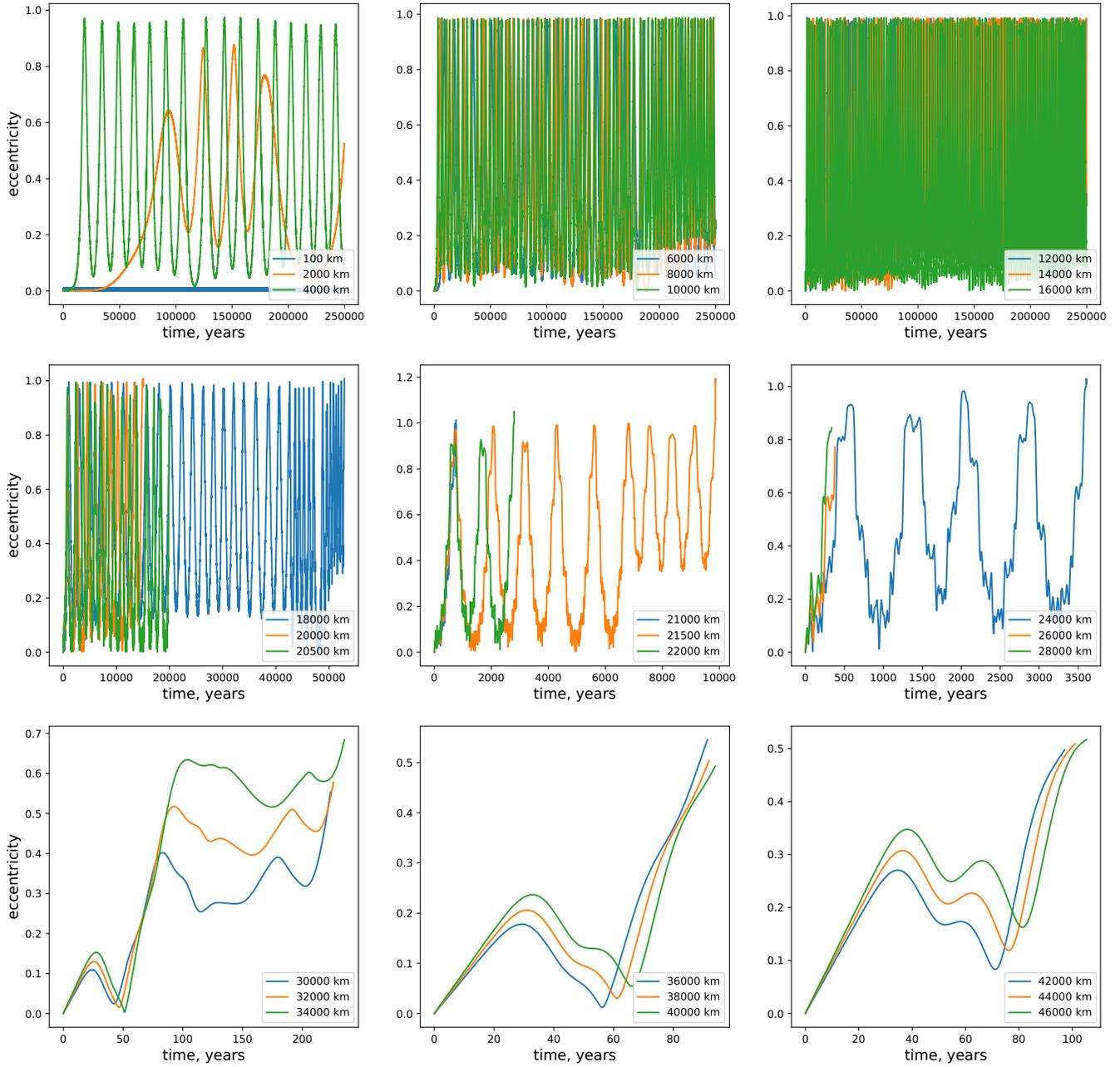}
\end{center}
\caption{The same as Fig.~\ref{fig4}, but for eccentricity.}
\label{fig5}
\end{figure}

\begin{figure}
\begin{center}
\includegraphics[width=1.0\linewidth]{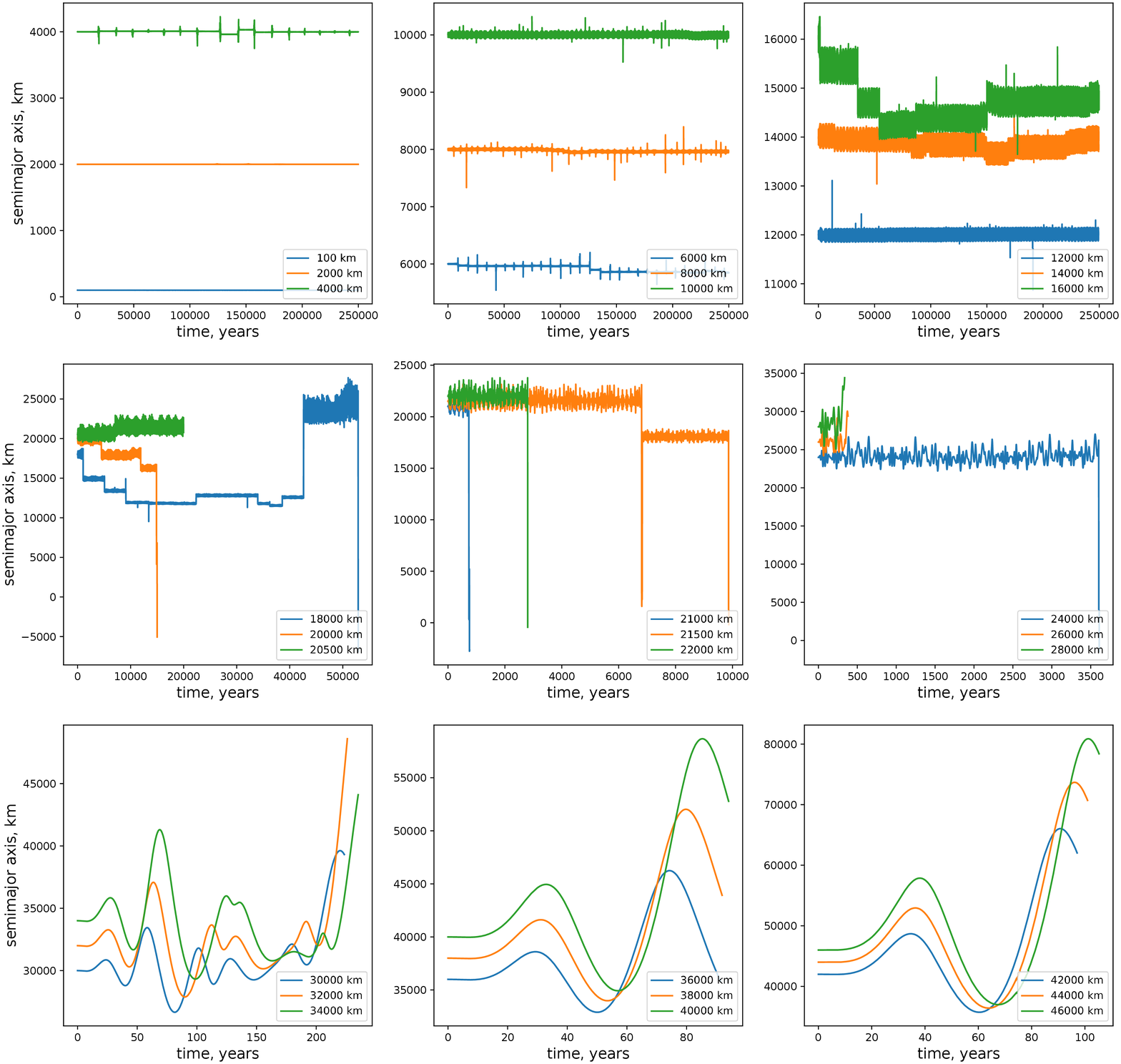}
\end{center}
\caption{The same as Fig.~\ref{fig4}, but for semimajor axis.
The semimajor axis excursions to negative values take place when
the orbit becomes hyperbolic, and the particle escapes.}
\label{fig6}
\end{figure}

\begin{figure}
\begin{center}
\includegraphics[width=1.0\linewidth]{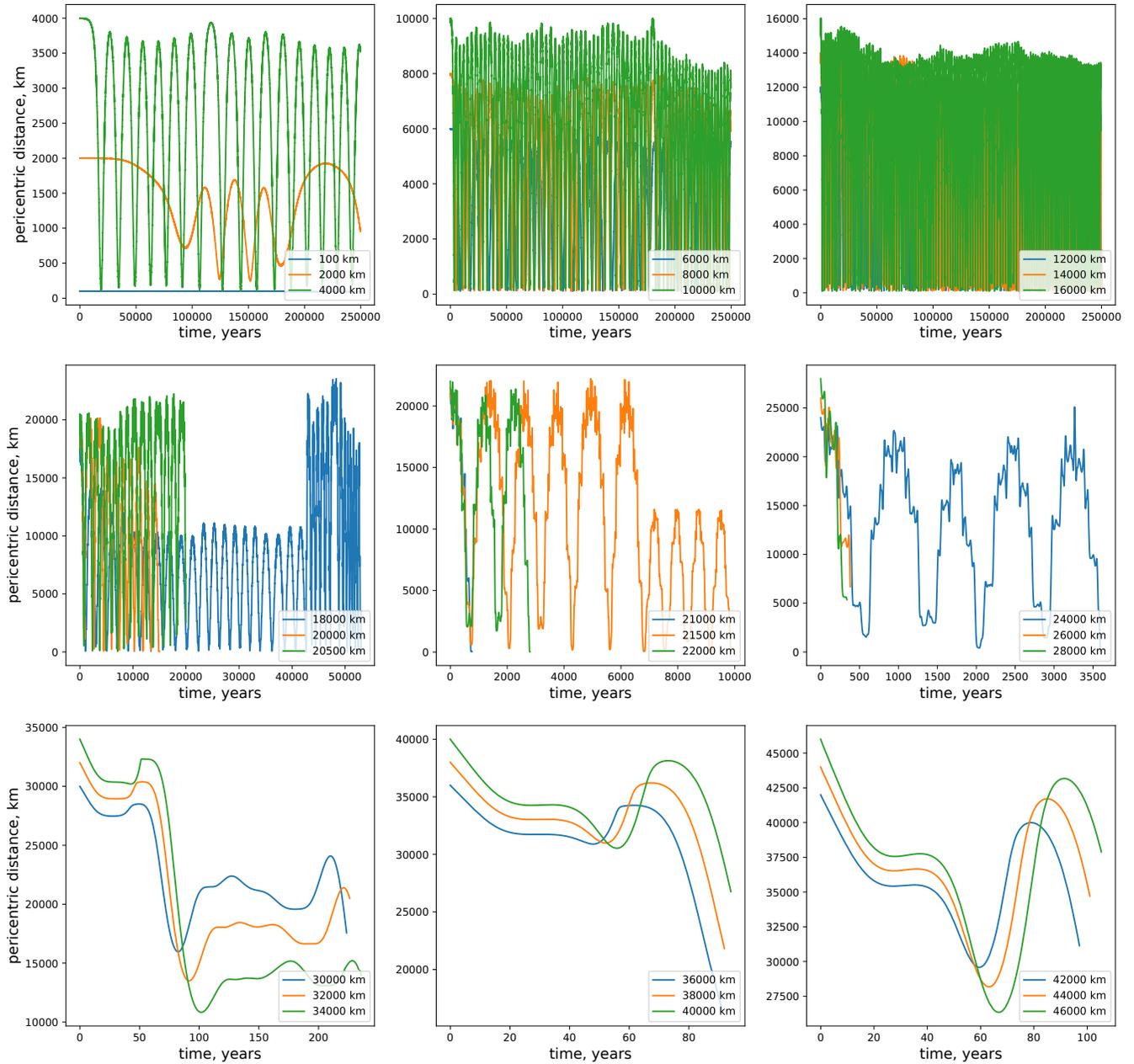}
\end{center}
\caption{The same as Fig.~\ref{fig4}, but for pericentric
distance.} \label{fig7}
\end{figure}

\begin{figure}
\begin{center}
\includegraphics[width=1.0\linewidth]{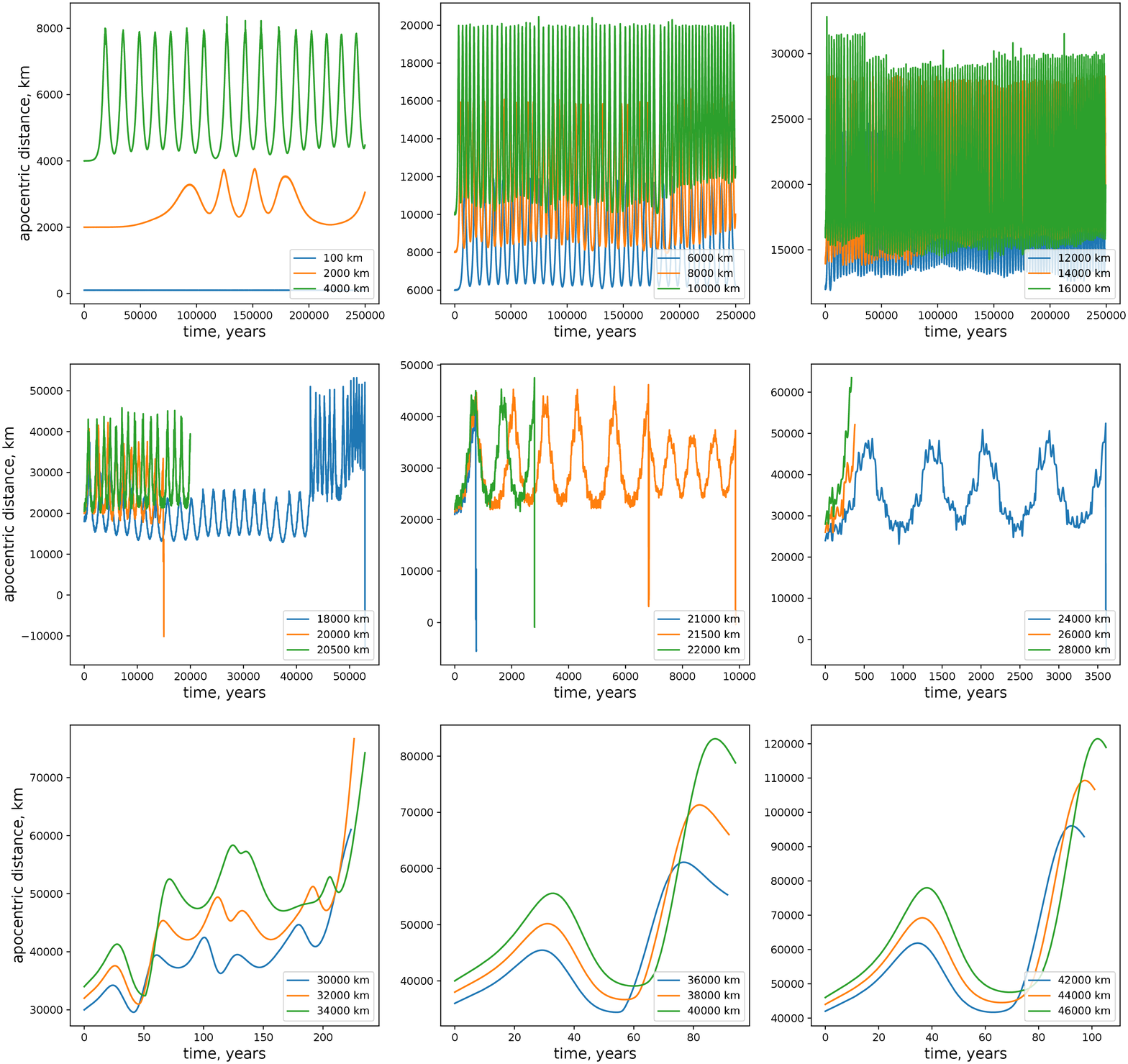}
\end{center}
\caption{The same as Fig.~\ref{fig4}, but for apocentric
distance.} \label{fig8}
\end{figure}

In Figs.~\ref{fig5}--\ref{fig8}, we present, for completeness of
the dynamical picture, the concurrent time evolution of
several other than inclination important orbital elements:
eccentricity, semimajor axis, pericentric and apocentric
distances. As in Fig.~\ref{fig4}, the curves of different
colours (blue, orange and green) correspond to three different
initial values of the semimajor axis.

From Figs.~\ref{fig4}--\ref{fig8} it follows that the
numerically observed timescales of cloudization of the outer
orbital zone are in accord with our theoretical estimates, based
on consideration of Solar perturbations, as provided by
Eqs.~\ref{T3D}; indeed, the orbital inclinations, in the
outer zone, rise macroscopically already on the scale of
Arrokoth's period of motion around the Sun.

In Figs.~\ref{fig4}--\ref{fig5} and \ref{fig7}--\ref{fig8}, the
Lidov--Kozai oscillations are readily recognizable, especially
clearly in the middle panels of these plots. What is more, as one
may deduce from the first (top left) panels, the Lidov--Kozai
effect is indeed suppressed at the orbital radii less than
$\sim$2000~km, in accord with our prediction made above using
Equation~(\ref{acrbc}): no definite LK-oscillations emerge if
$a_0$ is smaller than $\sim$2000~km.

According to \cite[section~3.2.2]{S17LKE}, the quantity
\begin{equation}
c_1 = (1 - e^2) \cos^2 i = \mathrm{const} , \label{C1cos}
\end{equation}
\noindent is conserved during the Lidov--Kozai oscillations, i.e.,
the vertical component of the angular momentum squared is
constant.
This relation implies that, if $0 \leq i \leq \pi/2$, then the
secular variations of $e$ and $i$ are coupled in anti-phase;
whereas, if $\pi/2 \leq i \leq \pi$, then the variations of $e$
and $i$ are coupled in phase.
The eccentricity is maximum at $i=0$, and the inclination is
maximum at $e=0$. If the initial inclination $i_0$ is greater than
a critical value ($\sim 40^\circ$; see \citealt{S17LKE} for review
and details), then the maximum eccentricity value is essentially
insensitive to the value of $e_0$ (if $e_0 \lesssim 0.1$) and can
be estimated by means of the formula

\begin{equation}
e_\mathrm{max} \approx \left( 1 - \frac{5}{3} \cos^2 i_0
\right)^{1/2} \label{emax}
\end{equation}

\noindent \citep{HTT97,IZM97AJ}.

In the Lidov--Kozai theory, the time-averaged semimajor axis is
constant with time. As one may see in Fig.~\ref{fig6}, the
semimajor axis is indeed approximately conserved on long intervals
of time, but sometimes it exhibits small jumps. The cause of these
fluctuations is as follows. According to Equation~(\ref{emax}), if
the initial $i_0 \sim 90^\circ$ (as in the given case), in the
course of LK-oscillations the eccentricity varies almost in the
whole interval from zero to unity; this means that the pericentric
distance periodically goes down to almost zero, and thus, from
time to time, the particle approaches the central rotating
dumbbell and receives, in accord with the Kepler map theory
\citep{S11NA,LSS17,LSS18Sch}, a kick in energy, which depends on
the approach distance and the dumbbell's orientation, when the
particle is at its orbital pericentre. The energy kick affects the
semimajor axis; therefore, the character of the following
LK-oscillations is modified, but is being sustained until the next
close approach to the CB takes place.

The fact that, in the course of the LK-oscillations, the
eccentricity varies here almost in the whole interval from zero to
unity means that the particle periodically approaches the
parabolic separatrix (the border between the elliptic and
hyperbolic types of motion) and thus, due to perturbations, may
easily become unbound. This explains the observed (in
Figs.~\ref{fig4}--\ref{fig8}) rapid removal of the low-mass matter
from the disk at the radii $\gtrsim$2000~km, where the
LK-oscillations are not suppressed.

On the other hand, as we have found out in
Section~\ref{sect_DmaA}, at radii $\lesssim$100~km the disk is
rapidly cleared because, if close enough to the rotating dumbbell,
the motion is chaotic. Therefore, solely the material that has
initial orbital radii in the interval from $\sim$100 to
$\gtrsim$2000~km may survive.

At the LK-resonance center, for a massless particle with orbital
period $P_\mathrm{orb}$ and semimajor axis $a$ (all other
notations are as adopted above) the period of LK-oscillations
can be rendered in the form

\begin{equation}
P_\mathrm{LK} \approx P_\mathrm{orb} \frac{m_\mathrm{A}}{m_0}
\left( \frac{a_1}{a} \right)^3 (1 - e_1^2)^{3/2} \label{PLK_0}
\end{equation}

\noindent \citep{MS79,HTT97,S17LKE}.
The periods of patterns in Figs.~\ref{fig4}--\ref{fig5} and
\ref{fig7}--\ref{fig8} are in accord with Equation~(\ref{PLK_0}).
E.g., according to Fig.~\ref{fig5}, the quasiperiodic eccentricity
variations, where they can be effectively identified, have the
time periods of $\sim$10000~yr (at $a_0=$4000, 6000 and 8000~km)
and $\sim$1000~yr (at $a_0=$21500, 22000 and 24000~km), and the
same approximate values are respectively produced by
Equation~(\ref{PLK_0}), as one may straightforwardly verify.

How principal is the circular approximation for Arrokoth's orbit
around the Sun? Setting $e_1 = 0.044$ (for Arrokoth's actual
elliptic orbit),
instead of $e_1 = 0$ in Equation~(\ref{PLK_0}), results in a
difference in $P_\mathrm{LK}$ of only $\sim$0.2\%; therefore,
taking into account Arrokoth's eccentricity is of no importance
for our estimates, which are made by the order of magnitude.
Besides, the smallness of this difference shows that the
eccentricity is as well essentially unimportant in numerical
integrations (of the kind presented in previous Sections), when
one is interested in the general dynamics character.

\begin{figure}
\begin{center}
\includegraphics[width=0.75\linewidth]{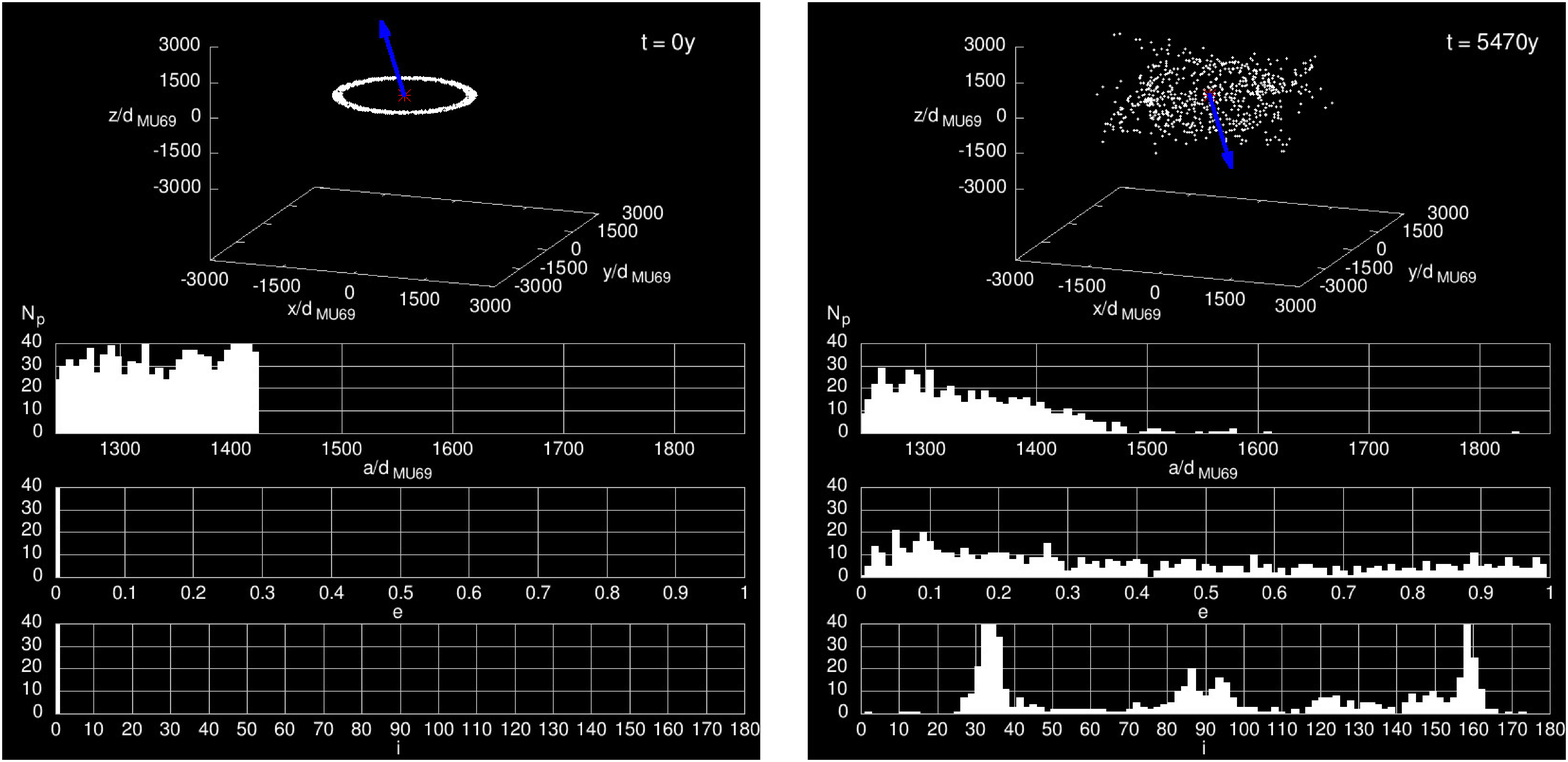}
\end{center}
\caption{Two snapshots (at time moments $t=0$ and $t=$5470~yr) of
the particles' cloud 3D evolution around Arrokoth. The initial (at
$t=0$) ring of the particles lies in the plane almost orthogonal
(inclination 99$^\circ$) to Arrokoth's spin axis. The mass
parameter $\mu=0.28$. The blue arrows give the direction to the
Sun during the simulation. The complete animation of the
evolution can be found at
https://search-data.ubfc.fr/FR-13002091000019-2022-08-05\_Dynamical-environments-of-Arrokoth-prior.html}
\label{fig9}
\end{figure}

As a graphical illustration of the ``cloudization''
phenomenon, an attached
animation\footnote{https://search-data.ubfc.fr/FR-13002091000019-2022-08-05\_Dynamical-environments-of-Arrokoth-prior.html}
and its two snapshots (Fig.~\ref{fig9}) demonstrates how an
initial planar ring of particles evolves, in the course of time,
into a complex 3D aggregate, arising due to Solar perturbations.
In this simulation, 1000 particles were initially placed in
circular orbits with initial radii ranging from $r_{\rm
min}=20000$~km to $r_{\rm max}=23000$~km. The simulation was
performed over 10000~yr; this is a typical time range as used
above to construct
Figs.~\ref{fig4}--\ref{fig8}.

Fig.~\ref{fig9}, right panel, exhibits the evolved 3D aggregate as
formed to the moment $t=5470$~yr (chosen here as a representative
one); the left panel demonstrates the initial particle
distribution. The evolved semimajor axes $a$ of the particles'
orbits are mostly distributed in the range 20000--26000~km;
however, in the course of evolution, some particles achieved
$a>R_\mathrm{H}$ and escaped (and were therefore removed from
the computation), in agreement with Figs.~\ref{fig4}--\ref{fig8}.
Furthermore, the eccentricity distribution rapidly becomes
uniform, whereas the inclination distribution shows broad peaks
whose number and locations vary with time. Analysis of these
complex spatial structures will be conducted elsewhere in further
studies.

\section{Conclusions}

In this article, we have considered dynamical environments of
(486958) Arrokoth, focusing on both their present state and their
long-term evolution, starting from the epoch of formation of the
object.

Both analytical (based on an upgraded Kepler-map formalism) and
numerical (based on massive simulations and construction of
stability diagrams) approaches to the problem have been used.

Our main conclusions are as following.

\begin{itemize}

\item In the 3D setting, the clearing process of the chaotic
circumbinary zone is practically instantaneous, as it is in the
planar case (explored in \citealt{RSL21}).

\item In the inner orbital zone (closer than 130~km to Arrokoth)
most of the particles (more than $\approx$60\%) that collide with
Arrokoth or escape its Hill sphere do it in $\sim$50~d. For
Arrokoth with its current rotation rate, the particles with
initial orbital radius less than 32~km collide with Arrokoth in
several days. For more distant ones, this takes up to
$\sim$10--100~yr.

\item The numerically observed timescales of cloudization of the
outer orbital zone are in accord with theoretical estimates, based
on consideration of Solar perturbations, as demonstrated and
discussed above in Section~\ref{sect_cloud}.

\item In the outer orbital zone, the particles that are initially
farther than $\sim$18000~km from Arrokoth cannot survive due to
Solar perturbations. The particles that are initially farther from
Arrokoth than $\sim$100~km and closer than $\sim$18000~km are
mostly stable.

\item The generic chaotization of Arrokoth's circumbinary debris
disk's inner zone and generic cloudization of the disk's
periphery, showed by us to be essential in the general 3D case,
naturally explains the current absence of any debris in its
vicinities.

\end{itemize}

\medskip

\noindent {\bf Acknowledgments.}
The authors are most thankful to Gustavo Madeira for valuable
remarks and comments. I.I.S. was supported in part by the
Russian Science Foundation, project 22-22-00046.

\medskip

\noindent {\bf Data availability.} The data underlying this
article will be shared on reasonable request to the corresponding
author.

\newpage

\begin{thebibliography}{}

\bibitem[Amarante \& Winter(2022)]{2022Ap&SS.367...38A}
Amarante A., Winter O.~C., 2022, Ap\&SS, 367, 38

\bibitem[Cheng et al.(2019)]{C19LPI}
Cheng A. F., et al., 2019, in: 50th Lunar and Planetary Science
Conference 2019, LPI Contrib. No. 2132, id.~3273

\bibitem[Everhart(1985)]{E85IAU}
Everhart E., 1985,
in: Carusi A., \& Valsecchi G.~B., eds, IAU Colloq. 83, Dynamics
of Comets: Their Origin and Evolution, v.~115, p.~185

\bibitem[Gladstone et al.(2019)]{G19LPI}
Gladstone G. R., et al., 2019, in: 50th Lunar and Planetary
Science Conference 2019, LPI Contrib. No. 2132, id.~2866

\bibitem[Holman, Touma, \& Tremaine(1997)]{HTT97}
Holman M., Touma J., \& Tremaine S., 1997, Nature, 386, 254

\bibitem[Innanen et al.(1997)]{IZM97AJ}
Innanen K. A., et al., 1997, Astron. J., 113, 1915

\bibitem[Jorda et al.(2016)]{J16Ic}
Jorda L., et al., 2016, Icarus, 277, 257

\bibitem[JPL database(2022)]{JPL22}
JPL database ``Solar System Dynamics'' (2022)
https://ssd.jpl.nasa.gov

\bibitem[Kammer et al.(2018)]{K18}
Kammer J., et al., 2018, Astron. J., 156, 72

\bibitem[Keane et al.(2022)]{K22JGR}
Keane J. T., et al., 2022, Journal of Geophysical Research:
Planets, 127 (6), e2021JE007068

\bibitem[Kozai(1962)]{K62AJ}
Kozai Y., 1962, Astron. J., 67, 591

\bibitem[Lages, Shepelyansky \& Shevchenko(2017)]{LSS17}
Lages J., Shepelyansky D. L., \& Shevchenko I. I., 2017, Astron.
J., 153, id.~272

\bibitem[Lages, Shepelyansky \& Shevchenko(2018)]{LSS18Sch}
Lages J., Shepelyansky D. L., \& Shevchenko I. I., 2018,
Scholarpedia, 13(2):33238

\bibitem[Lages, Shevchenko \& Rollin(2018)]{LSR18}
Lages J., Shevchenko I. I., \& Rollin G., 2018, Icarus, 307, 391

\bibitem[Lidov(1961)]{L61I}
Lidov M. L., 1961, Iskusstviennye Sputniki Zemli (Artificial
Satellites of the Earth), 8, 5 (in Russian)

\bibitem[Lidov(1962)]{L62PSS}
Lidov M. L., 1962, Planet. Space Sci., 9, 719

\bibitem[Madeira et al.(2022)]{madeira22}
Madeira G., Giuliatti Winter S. M., Ribeiro T., Winter O. C.,
2022, MNRAS, 510, 1450

\bibitem[Malyshkin \& Tremaine(1999)]{MT99}
Malyshkin L., \& Tremaine S., 1999, Icarus, 142, 341

\bibitem[Mazeh \& Shaham(1979)]{MS79}
Mazeh T., \& Shaham J., 1979, Astron. Astrophys., 77, 145

\bibitem[McKinnon et al.(2019)]{MK19LPI}
McKinnon W. B., et al., 2019, in: 50th Lunar and Planetary Science
Conference 2019, LPI Contrib. No. 2132, id.~2767

\bibitem[Meiss(1992)]{Meiss92}
Meiss J. D., 1992, Rev. Mod. Phys., 64, 795

\bibitem[Michaely et al.(2017)]{MPG17}
Michaely E., Perets H. B., \& Grishin E., 2017, Astrophys. J.,
836, 27

\bibitem[Morbidelli(2002)]{Morbi02}
Morbidelli A., 2002, Modern Celestial Mechanics. Aspects of Solar
System Dynamics (Taylor and Francis, London)

\bibitem[Mysen, Olsen \& Aksnes(2006)]{mysen06}
Mysen E., Olsen \O., Aksnes K., 2006, Planet. Space Sci., 54, 750

\bibitem[Nesvorn\'y et al.(2018)]{N18AJ}
Nesvorn\'y D., et al., 2018, Astron. J., 155, 246

\bibitem[Parker et al.(2017)]{P17}
Parker A. H., et al., 2017,
in: 49th Meeting of the AAS Division for Planetary Sciences,
abstract~504.04

\bibitem[Porter et al.(2019)]{P19EPSC}
Porter S., et al., 2019,
in: EPSC-DPS Joint Meeting 2019, v.~2019, p. EPSC--DPS2019--311,
Sept. 2019

\bibitem[Protopapa et al.(2019)]{P19LPI}
Protopapa W. M., et al., 2019, in: 50th Lunar and Planetary
Science Conference 2019, LPI Contrib. No. 2132, id.~2732

\bibitem[Rollin, Shevchenko \& Lages(2021)]{RSL21}
Rollin G., Shevchenko I. I., \& Lages J., 2021, Icarus, 357,
id.~114178

\bibitem[Shao \& Lu(2000)]{SL00JGR}
Shao Y., \& Lu H., 2000, J. Geophys. Res., 105, 22437

\bibitem[Shevchenko(2011)]{S11NA}
Shevchenko I. I., 2011, New Astronomy, 16, 94

\bibitem[Shevchenko(2017)]{S17LKE}
Shevchenko I. I., 2017, The Lidov--Kozai Effect -- Applications in
Exoplanet Research and Dynamical Astronomy (Springer International
Publishing Switzerland AG)

\bibitem[Spencer et al.(2019)]{SSL19LPI}
Spencer J. R., et al., 2019, in: 50th Lunar and Planetary Science
Conference 2019, LPI Contrib. No. 2132, id.~2737

\bibitem[Spencer et al.(2020)]{SSM20Sci}
Spencer J. R., et al., 2020,
Science, 367, No.~6481, p.~eaay3999

\bibitem[Stern(2017)]{S17}
Stern A., 2017, The PI's Perspective: The Heroes of the DSN and
the `Summer of MU69'. (August~8, 2017.) Available at:
http://pluto.jhuapl.edu/News-Center/

\bibitem[Stern et al.(2019a)]{S19LPI}
Stern S. A., et al., 2019a, in: 50th Lunar and Planetary Science
Conference 2019, LPI Contrib. No. 2132, id.~1742

\bibitem[Stern et al.(2019b)]{Stern19Sci}
Stern S. A., et al., 2019b,
Science, 364, eaaw9771

\bibitem[Thirouin et al.(2014)]{TNO14}
Thirouin A., Noll K. S., Ortiz J. L., \& Morales N., 2014, Astron.
Astrophys., 569, A3

\bibitem[Thomas et al.(2013)]{T13Ic}
Thomas P., et al., 2013, Icarus, 222, 550

\bibitem[Thomas et al.(2015)]{T15AA}
Thomas N., et al., 2015, Astron. Astrophys., 583, A17

\bibitem[Tremaine(1993)]{T93}
Tremaine S., 1993, in: Phillips J. A., Thorsett J. E., \& Kulkarni
S. R., eds, Planets Around Pulsars, ASP Conf. Series, 36, 335

\bibitem[Umurhan et al.(2019)]{U19LPI}
Umurhan O. M., et al., 2019, in: 50th Lunar and Planetary Science
Conference 2019, LPI Contrib. No. 2132, id.~2809

\end{thebibliography}
\end{document}